\documentclass[12pt,a4paper,dvips]{article}
\usepackage{epsfig,wrapfig,times,mathptm} 
\renewcommand{\section}[1]{\vspace{6pt} \noindent\mbox{#1} \newline
\noindent}
\renewcommand{\subsection}[1]{\vspace{6pt}
\noindent\mbox{\underline{#1}} 
\newline \noindent}
\renewcommand{\subsubsection}[1]{\vspace{6pt}
\noindent\mbox{\underline{#1}}
\noindent}

\newfont{\sansb}{cmssbx10}
\newfont{\sans}{cmss10}

\begin{document}
{\small OG 10.3.26 \vspace{-24pt}\\}     
{\center \LARGE VERITAS: VERY ENERGETIC RADIATION IMAGING TELESCOPE ARRAY
SYSTEM
\vspace{6pt}\\}
T.C.Weekes$^1$, C.Akerlof$^{12}$, S.Biller$^{16}$,
A.C.Breslin$^{15}$, J.H.Buckley$^1$, D.A.Carter-Lewis$^6$,
M.Catanese$^6$, M.F.Cawley$^{15}$, B.Dingus$^{13}$, G.G.Fazio$^1$,
D.J.Fegan$^{15}$, J.Finley$^9$, G.Fishman$^7$, J.Gaidos$^9$,
G.H.Gillanders$^{15}$, P.Gorham$^3$, J.E.Grindlay$^1$,
A.M.Hillas$^{17}$, J.Huchra$^1$,  P.Kaaret$^4$, M.Kertzman$^5$,
D.Kieda$^{13}$, F.Krennrich$^6$, R.C.Lamb$^3$, M.J.Lang$^{15}$,
A.P.Marscher$^2$, S.Matz$^8$, T.McKay$^{12}$, D.M\" uller$^{11}$,
R.Ong$^{11}$, W.Purcell$^8$, J.Rose$^{17}$, G.Sembroski$^9$,
F.D.Seward$^1$, P.Slane$^1$,  S.Swordy$^{11}$, T.T\" umer$^{10}$,
M.Ulmer$^8$, M.Urban$^{14}$, and B.J.Wilkes$^1$ 
\vspace{6pt}\\
{\it$^1$Smithsonian Astrophysical Observatory, Cambridge, MA
02138\\ 
$^2$Boston University, Boston, MA 02215\\
$^3$California Institute of Technology, Pasadena, CA 91125\\
$^4$Columbia University, New York, NY 10027\\
$^5$De Pauw University, Greencastle, IN 46135\\
$^6$Iowa State University, Ames, IA 50011\\
$^7$NASA, Marshall S.F.C., Huntsville, AL 35812\\
$^8$Northwestern University, Evanston, IL 60208\\
$^9$Purdue University, West Lafayette, IN 47907\\
$^{10}$University of California, Riverside, CA 92521\\
$^{11}$University of Chicago, Chicago, IL 60637\\
$^{12}$University of Michigan, Ann Arbor, MI 48109\\
$^{13}$University of Utah, Salt Lake City, UT 84112\\
$^{14}$Ecole Polytechnique, Palaiseau, France\\
$^{15}$National University of Ireland, Dublin, Ireland \\
$^{16}$Oxford University, Oxford, U.K. \\
$^{17}$University of Leeds, Leeds, LS2 9JT, U.K. 
\vspace{-12pt}\\}
{\center ABSTRACT\\}
A next generation atmospheric Cherenkov observatory is described
based on the Whipple
Observatory $\gamma$-ray telescope.  A total of nine such imaging
telescopes will be deployed in an array that will permit the
maximum versatility and give high sensitivity in the 50 GeV - 50
TeV band (with maximum sensitivity from 100 GeV to 10 TeV).

\setlength{\parindent}{1cm}
\section{INTRODUCTION}
Recent successful detections in Very High Energy 
(VHE) $\gamma$-ray
astronomy using the atmospheric Cherenkov technique have triggered
a spate of projects aimed at extending and improving the detection
technique. Although these projects (at various stages of conceptual
design, detailed planning or actual construction) have radically
different experimental approaches, they have important features in
common:
(i) all are based on the belief that the scientific benefits gained
by an increase in sensitivity justify a major effort;
(ii) all are agreed that major improvements in sensitivity are
physically possible and technically straightforward; and
(iii) by the standards normally used in this field all the projects
are expensive (in the \$3M to \$15M range).  

State-of-the-art imaging Atmospheric Cherenkov Telescopes
(ACTs) are best represented by:  the Whipple
Collaboration telescope in southern Arizona [Reynolds et al. 1993],
the French CAT telescope in the French Pyrenees [Punch 1995], the
Armenian-German-Spanish HEGRA telescope array in the Canary Islands
[Petry et al. 1995], the Durham telescope in Narrabri, Australia
[Chadwick et al. 1995], the Australian-Japanese telescopes at
Woomera, Australia [Tanimori et al. 1995], the Russian SHALON-
ALATOO project at Tien-Shan [Sinitsyna 1995], the Crimean
Astrophysical Observatory telescopes [Zyskin et al. 1995], the
Indian TACTIC Array at Mount Abu, India [Bhat et al. 1993] and the
Japanese Telescope Array in Utah, USA [Hayashida et al. 1996].

The Whipple Observatory $\gamma$-ray telescope consists of a
10m diameter optical reflector focussed onto an array of PMTs.
The energy threshold is $\sim$300 GeV
and the telescope 
routinely obtains a 5$\sigma$ excess from the Crab Nebula
in one-half hour of on-source observations. In 1996 the 109 pixel
camera was expanded to 151 pixels and eventually it will have 541
pixels [Lamb et al. 1995].
Although the 10m reflector was built in 1968 and the first imaging
camera installed in 1983, this is still the prototype device for
atmospheric Cherenkov imaging systems.

The features of VHE ACTs that can be improved include: (a) energy
threshold; (b) flux sensitivity; (c) energy resolution; (d) angular
resolution; (e) field of view. Of these, energy threshold is the
easiest to achieve since energy threshold scales as (mirror
area)$^{-1}$. The proposed detector, VERITAS, would make
improvements in all these parameters. 

\section{VERITAS}
\begin{figure}
\centerline{\epsfig{file=
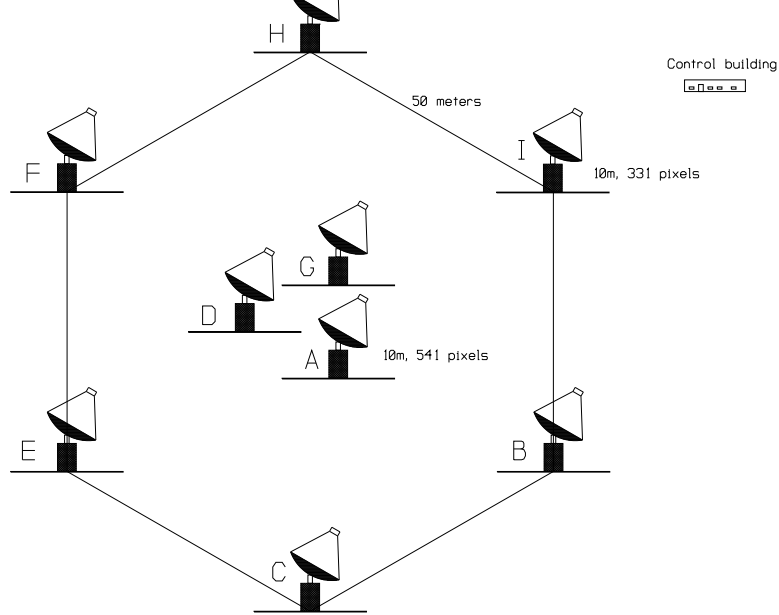,
height=4in,angle=270.}}
\caption{Layout of telescopes in VERITAS.}
\label{athena-fig}
\end{figure}

The philosophy underlying VERITAS comes from 30 years of
development of ACTs at the Whipple Observatory [Weekes 1996]; the
objective is to build a VHE $\gamma$-ray observatory which will
have a useful lifetime well into the next century. The initial aim
is to have the maximum sensitivity in the 100GeV-10TeV range but to
have significant sensitivity down to 50 GeV (and lower as new
technology photo-detectors become available) and as high as 50TeV
(using the low elevation technique [Krennrich 
et al. 1997]). The detection
technique will be the so-called ``imaging" atmospheric Cherenkov
technique which was originally demonstrated at the Whipple
Observatory but is now under considerable development at a number
of centers. The strawman design for VERITAS will have the basic telescopes  modelled on the Whipple
10m telescopes with wide field cameras of 331 to 541 pixels. The
array will consist of nine such telescopes, all capable of
independent or coincident operation. The telescope layout will be
as shown in Figure~\ref{athena-fig}. A somewhat similar array 
(HESS) has
been proposed by the Heidelberg group (F.Aharonian, private
communication); it would have 16 telescopes with wider separations
than envisaged here and would probably be located in Spain.

The preferred location of VERITAS is a flat area at the Whipple
Observatory Basecamp (elevation 1.3km) where there is ample space
for development as well as easy access to roads, power, etc.
An alternative site, with somewhat less area, is also under consideration;
this is at the 2.15km level of Mt. Hopkins, just below the present site of   
the Whipple telescope.
Southern Arizona has been shown to be an excellent site for these
kinds of astronomical investigation with an impressive record of
clear nights. These dark sites is not environmentally sensitive nor is
there the potential for conflict with other astronomical
activities.

\begin{figure}
\vspace*{-0.55in}
\centerline{\epsfig{file=
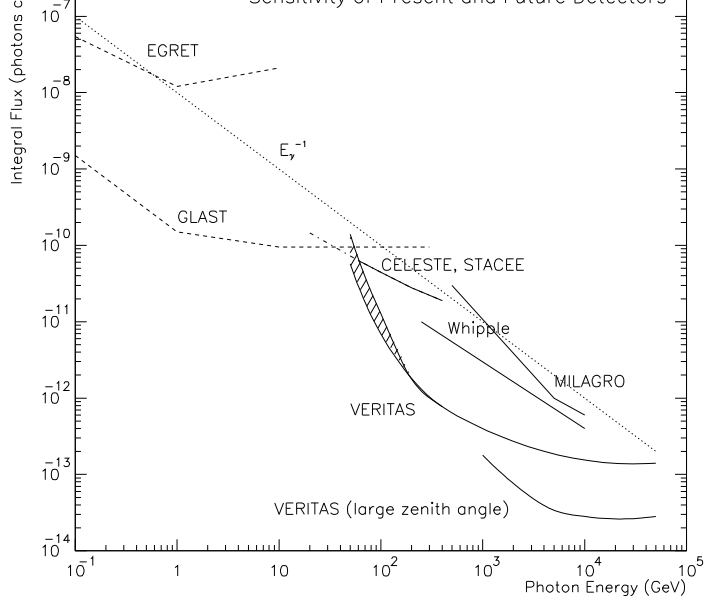,
height=4.0in,angle=0.}}
\vspace*{-0.2in}
\caption{Predicted sensitivity of VERITAS. Also shown are the known
sensitivities of EGRET and Whipple and the predicted sensitivities
of the MILAGRO, STACEE, CELESTE and GLAST experiments, all of which
are at various stages of planning and construction. The exposure
for VERITAS, STACEE and CELESTE is 50 hours. The exposure for
EGRET, GLAST and MILAGRO is one year of sky survey operation.}
\label{sensit-fig}
\end{figure}

The parameters of the array are chosen to give the optimum flux
sensitivity in the 100GeV-10TeV range which has proven to be rich
in scientific returns. The predicted flux sensitivity is shown in
Figure~\ref{sensit-fig}; it is seen to be a factor of ten better
than any other detector in this range. In these two decades of
energy the major background comes from hadron-initiated air showers
for which successful identification methods have been developed. At
the lower end single muons become the major background but these
can be removed by the coincident requirement in the separated
telescopes Also
at lower energies, the cosmic electron background constitutes an
irreducible isotropic background. Over these two decades of energy
the angular and energy resolutions will be pushed to their limits
(0.05$^o$ and 8\% respectively).

\section{CONCLUSIONS}
There are a number of alternative projects designed to increase the
sensitivity of telescopes in the 10GeV-10TeV range. These include
the solar farm projects (STACEE in the USA [Ong et al. 1995] and
CELESTE in France [Quebert et al. 1995]), the single dish approach
(MAGIC in Germany [Mirzoyan 1997]), the large water Cherenkov air
shower detector (MILAGRO in New Mexico, USA [Yodh 1996]), the next
generation space telescope (GLAST which will not be launched before
2004 [Gehrels et al. 1997]). All of these have merit in particular
areas; the relative merits of these projects are compared with
VERITAS for different experimental parameters in Table 1. In this
table the ranking system is *** = excellent; ** = very good: * =
good and blank = not good. 
Timiliness is defined as how quickly will the project come to fruition and cost is in inverse proportion to the total expenditure. Although the *'s are
toted, a better representation might be * per dollar.
Since both the choice of parameters and
the ranking are assigned by the lead author it is not surprising
that VERITAS compares favorably with all the other projects.

\begin{table}[thb]
 \begin{center}
 \caption{\em {Comparison of Proposed Projects}}
  \begin{tabular}{|c|c|c|c|c|c|} \hline

Concept & Solar Farm & Single Dish & Array & Particle & Space  \\
\hline
Project & STACEE & MAGIC & VERITAS & MILAGRO  & GLAST \\
Energy Threshold & *** & ** & * & * & *** \\
Dynamic Range & * & ** & ** & ** & *** \\
Flux Sensitivity & * & ** & *** & * & *** \\
Energy Resolution & * & * & *** & * & ** \\
Angular Resolution & * & ** & *** & * & ** \\
Field of View &  & ** & *** & *** & *** \\
Cost & *** & ** & ** & ** &   \\    
Timeliness & *** & ** & ** & *** & * \\ 
TOTAL (*) & 13 & 15 & 19 & 14 & 17 \\
\hline
  \end{tabular}
 \end{center} 
\end{table}

\vspace*{12pt}
\section{ACKNOWLEDGMENTS}
This work is supported by a grant from the U.S. Department of
Energy.

\section{REFERENCES}
\setlength{\parindent}{-5mm}
\begin{list}{}{\topsep 0pt \partopsep 0pt \itemsep 0pt \leftmargin
5mm
\parsep 0pt \itemindent -5mm}
\vspace{-15pt}

\item Bhat, C.L. et al., Proc. of Workshop on Major Atmospheric
Cherenkov Detectors-II, Calgary, Canada, Ed. R.C.Lamb, Publ. Iowa
State Univ., 101 (1993).

\item Chadwick, P.M. et al., Workshop on Gamma-Ray
Astrophysics, Padua, Ed. M.Cresti, 301 (1995).

\item Gehrels, N. et al., Proc. 4th Compton Symposium, Willamsburg,
Virginia (in press) (1997).

\item Hayashida, N. et al., Proc. Int. Symposium on Extremely High
Energy Cosmic Rays, Tanashi, Japan Ed. M.Nagano, 205 (1996).

\item Krennrich, F. et al., ApJ, 481, 758  (1997).

\item Lamb, R.C. et al., Proc. ICRC (Rome), 2, 491 (1995).

\item Mirzoyan, R., Proc. Int. Symposium on Extremely High Energy
Cosmic Rays, Tanashi, Japan Ed. M.Nagano, 329 (1996).

\item Ong, R.A. et al., Proc. Workshop on Gamma-Ray
Astrophysics, Padua, Ed. M.Cresti, 241 (1995).

\item Petry, D. et al., Proc. Workshop on Gamma-Ray
Astrophysics, Padua, Ed. M.Cresti, 141 (1995).

\item Punch, M., Proc. Workshop on Gamma-
Ray Astrophysics, Padua, Ed. M.Cresti, 356 (1995).

\item Quebert, J. et al., Proc. Workshop on Gamma-
Ray Astrophysics, Padua, Ed. M.Cresti, 248 (1995).

\item Reynolds, P.T. et al., ApJ. 404, 206 (1993).

\item Sinitsyna, V.G., Proc. Workshop on Gamma-
Ray Astrophysics, Padua, Ed. M.Cresti, 133 (1995).

\item Tanimori, T. et al., Proc. Workshop on Gamma-
Ray Astrophysics, Padua, Ed. M.Cresti, 316 (1995).

\item Weekes, T.C., Space Sci. Rev. 75, 1 (1996).

\item Yodh, G.B., Proc. Int. Symposium on Extremely High Energy
Cosmic Rays, Tanashi, Japan Ed. M.Nagano, 341 (1996).

\item Zyskin, Yu.L. et al., J.Phys. G. 20, 1851 (1994).
 
\end{list}

\end{document}